\documentclass[12pt]{article}
\usepackage{amsmath}
\usepackage{amssymb}
\usepackage{graphicx,psfrag,epsf}
\usepackage{enumerate}
\usepackage{natbib}
\usepackage{subcaption}
\usepackage{url} 
\usepackage[hidelinks]{hyperref}
\usepackage{float}

\newcommand{\blind}{0}

\newcommand{\bcx}{{\bf X}}

\newcommand{\bcz}{{\bf Z}}

\newcommand{\bc}{{\bf c}}
\newcommand{\bx}{{\bf x}}

\newcommand{\bz}{{\bf z}}

\newcommand{\bmu}{{\boldsymbol\mu}}
\newcommand{\bfphi}{{\boldsymbol\varphi}}
\newcommand{\bfeta}{{\boldsymbol\eta}}
\newcommand{\beps}{{\boldsymbol\epsilon}}
\newcommand{\btheta}{{\boldsymbol\theta}}

\newcommand{\bfsigma}{{\boldsymbol\sigma}}

\newcommand{\R}{\mathbb{R}}

\begin{document}

\def\spacingset#1{\renewcommand{\baselinestretch}%
{#1}\small\normalsize} \spacingset{1}

\newcommand{\printtitle}{\bf A statistical approach to latent dynamic modeling with differential equations}


\if0\blind
{
  \title{\printtitle}
  \author{Maren Hackenberg\thanks{
    This work was funded by the Deutsche Forschungsgemeinschaft (DFG, German Research Foundation) -- Project-ID 322977937 -- GRK 2344 (MH) and Project-ID 499552394 -- SFB 1597 (HB, CK, and MH).
    AP was supported by the Berta-Ottenstein clinician scientist program of the University of Freiburg. 
    Biogen and Novartis provide financial support for the SMArtCARE registry. \hspace{.2cm}}\\
    Institute of Medical Biometry and Statistics, \\ Faculty of Medicine and Medical Center, University of Freiburg \\ 
    and \\
    Astrid Pechmann \\
    Department of Neuropediatrics and Muscle Disorders, \\ Faculty of Medicine and Medical Center, University of Freiburg \\ 
    and \\ 
    Clemens Kreutz \\ 
    Institute of Medical Biometry and Statistics, \\ Faculty of Medicine and Medical Center, University of Freiburg; \\
    Freiburg Center for Data Analysis and Modeling, University of Freiburg; \\ Centre for Integrative Biological Signaling Studies (CIBSS), University of Freiburg \\
    and \\
    Janbernd Kirschner \\
    Department of Neuropediatrics and Muscle Disorders, \\ Faculty of Medicine and Medical Center, University of Freiburg \\
    and \\
    Harald Binder \\
    Institute of Medical Biometry and Statistics, \\ Faculty of Medicine and Medical Center, University of Freiburg; \\
    Freiburg Center for Data Analysis and Modeling, University of Freiburg; \\ Centre for Integrative Biological Signaling Studies (CIBSS), University of Freiburg \\
    }
  \maketitle
} \fi

\if1\blind
{
  \bigskip
  \bigskip
  \bigskip
  \begin{center}
    {\LARGE \printtitle}
\end{center}
  \medskip
} \fi

\newpage

\begin{abstract}
    Ordinary differential equations (ODEs) provide mechanistic models of temporally local changes of processes, where parameters can be informed by external knowledge. For statistical modeling of longitudinal cohort data, the use of ODEs is so far limited compared to regression-based global function fitting approaches, yet modeling local changes based on an individual’s current status with ODEs could also be attractive in a clinical cohort setting. 
    This is potentially due to a larger number of variables to be modeled and a higher noise level, as the shape of an ODE solution strongly depends on the initial value.
    To address this, we propose to use each observation as the initial value to obtain multiple local ODE solutions, supporting subsequent prediction starting from arbitrary time points, and build a combined estimator. We use neural networks for obtaining a low-dimensional latent space for dynamic modeling with many variables, and for obtaining individual-specific ODE parameters from baseline variables. Differentiable programming allows for simultaneously fitting dynamic models and a latent space. 
    We illustrate the approach in an application with spinal muscular atrophy patients and contrast modeling of local changes in health status to global fitting via regression, highlighting how different application settings might demand different modeling strategies.
\end{abstract}

\noindent%
\vfill


\newpage
\spacingset{1.45} 
\section{Introduction}
\label{sec:intro}

Different modeling communities have developed distinct approaches to describing temporal processes that underlie longitudinal data, such as the progression of a disease over time.  
In statistics, analysis of longitudinal cohort data is typically based on regression techniques \citep{Fitzmaurice2008longitudinal, BanGanCha2011}, which provide a global model that is a good fit on average over the entire observed time course, e.g., corresponding to the average course of a typical individual from a given cohort. 
Yet, sometimes the future trajectory of a specific individual, given the current status, might be of interest. For example, a clinical practitioner might want to predict how a patient's health status will develop until the next follow-up visit given the current status. This corresponds to focusing on relative changes and a local perspective on the dynamics, as in ordinary differential equations (ODEs). The latter are commonly used, e.g., in systems biology \citep{BruWes2007, Banga2008, Distefano2015dynamic}, but less established in the statistics community. ODEs describe a small set of quantities by carefully specified relations, typically based on domain knowledge. 

In some settings, information about the dynamics are known beforehand. For example, in a systems biology context, kinetic parameters of biochemical reactions might be known from thermodynamics and initial conditions depend on the experimental context. In a clinical cohort setting, a typical disease progression might be determined by patients' characteristics at baseline, e.g., the age at diagnosis. 
In such settings, the observation at the first time point could be used as an initial value, which will then strongly influence the shape of the solution. This might be problematic particularly for noisy longitudinal cohort data \citep{Xue2010sieve, Verbeke2014}. In addition, specifying the relations of a larger number of variables with ODEs is challenging, as sufficient domain knowledge may not be available and the resulting complex systems may be difficult to solve numerically.

To still enable modeling with ODEs for describing individual local changes in such a setting, yet reduce the emphasis on the initial value and increase the robustness to noise, we propose a statistically inspired approach based on ODEs. In addition, we propose to integrate neural network techniques for dimension reduction of a larger number of variables, and for inferring individual ODE parameters from the characteristics of an individual. 

Neural networks allow for building more flexible approaches, e.g., for modeling with unobserved quantities in a latent representation. Such combinations of data-driven modeling based on neural networks with knowledge-driven modeling based on ODEs are explored, e.g., in the framework of universal differential equations \citep{Rackauckas2020} based on the idea of neural ODEs \citep{Chen2018, Rubanova2019}, and have also been investigated in biomedical applications, e.g., for intervention modeling \citep{Gwak2020}, survival analysis \citep{Moon2022}, or modeling of psychological resilience \citep{Koeber2022individualizing, Koeber2022deep}. 
Such a combination of modeling components is facilitated by differentiable programming, a paradigm that allows for joint optimization of different model components via automatic differentiation \citep{Baydin2017}. While originally proposed to bridge scientific computing and machine learning \citep{Innes2019_Zygote_dP}, we have argued that such integrative approaches can also be applied to statistical modeling \citep{Hackenberg2022Using}. 

The proposed ODE-based approach shares conceptual similarities with physics-informed neural networks (PINNs) \citep{Cuomo2022}, which integrate physical laws and data-driven methods to model complex systems. Similar to our approach, PINNs combine mechanistic insights with data-driven modeling. 
However, they are typically designed for densely sampled continuous systems, while our approach is tailored to sparse, irregularly sampled multivariate longitudinal data, typical for biomedical applications. 

In settings with more classical statistical models, larger numbers of variables have been reduced to a single summary score \citep{BanGanCha2011, Fitzmaurice2008longitudinal} and subsequently analyzed with univariate methods. Yet,  this means that dimension reduction and longitudinal modeling are performed separately. 
In some approaches, the covariance structure between outcomes has been included explicitly or implicitly into longitudinal regression \citep{Fan2007analysis, BanGanCha2011, Fitzmaurice2008longitudinal}. For example, functional principal component analysis on the covariance surface can be used to identify a smooth joint trajectory and individual deviations \citep{Nolan2023}, or mixed-effects models can be adapted, e.g., for explicitly modeling intra-individual variability \citep{Palma2023}, or for capturing individual trajectories via spatiotemporal transformations \citep{Schiratti2017}.
In contrast, auto-regressive approaches dynamically adapt to incoming data by regressing the observed variable on its past \citep{Montfort2018continuous}. For modeling in a latent space, long-short term memory (LSTM) models infer a latent state from observed data using neural networks, which is then iteratively updated \citep{Hochreiter1997long, Graves2012long}. Yet, these models operate in discrete time and typically rely on a fixed time grid of observations. While there are some examples of continuous-time modeling in autoregressive models \citep{Wang2013} and statistical applications \citep{Ramsay2017dynamic, Haan2017discrete, Mews2022}, often the data is too sparse and noisy to allow for fitting ODEs \citep{Montfort2018continuous}, resulting in limited use in the statistics community so far. Yet, we have shown in our own work that model fitting is feasible even with very limited information, such as only few time points \citep{HacHarPfa2022}. While this entailed some over-simplification, such as having to designate a specific time point for the ODE initial values, the overall modeling strategy shows considerable promise.

Conversely, statistical function fitting techniques have been proposed to estimate ODE parameters from an observed time series in systems modeling \citep{Liang2008parameter, Gabor2015, Ramsay2017dynamic}, e.g., using maximum likelihood estimation and uncertainty quantification via the profile likelihood \citep{Raue2009, Kreutz2012}. Parameter estimation is then based on longitudinal information, i.e., corresponds to the function fitting approach of regression techniques. 
Initial values can also be considered as additional parameters to account for observation noise and can then be estimated jointly with the parameters specifying the dynamics. Yet, such models are typically used in settings with a smaller number of variables and not designed for modeling in a latent space. In addition, it could be attractive to obtain the ODE parameters not based on function fitting, but based on external information from the individuals, to model each time series individually. This might be particularly important when modeling data with heterogeneous individual dynamics. 

We therefore propose an approach where ODE parameters are not obtained via function fitting, but from baseline characteristics of individuals via a neural network. For modeling with a larger number of variables, we fit the dynamic model in a low-dimensional latent space learned by a neural network, specifically a variational autoencoder (VAE) \citep{Kingma2014}, to reflect the assumption of a lower-dimensional underlying dynamic process driving the observed quantities. 
To prevent the propagation of the noise through the system when using noisy observations as initial values for ODEs, we propose to solve multiple ODEs, using each observed value as the initial value and averaging the solutions using time-dependent inverse-variance weights. This averaging process mitigates the effect of noise and reduces the dependence on the first initial value, thus improving robustness. In particular, this remedies potential over-simplification of the approach that we had presented in \cite{HacHarPfa2022}, and therefore provides a better foundation for subsequently obtaining predictions starting at arbitrary time points.

Before explaining the proposed approach in detail, we illustrate conceptual differences between global function fitting approaches and ODE-based local approaches, motivated by an application example from the SMArtCARE rare disease registry \citep{Pechmann2019}, which highlights how different research questions demand different modeling perspectives, thus more generally motivating our proposed approach.  
After providing an in-depth description of the approach, we use the SMArtCARE data and a corresponding simulation design to contrast the resulting individual-specific local models of changes in dynamics at any point in time with a global function fitting approach.
Finally, we discuss limitations and the more general potential of fusing dynamic modeling approaches across communities.

\section{An illustration of local and global perspectives in a disease registry application}
\label{sec:illustration}

We illustrate the individual-level temporally local perspective on dynamic modeling as compared to temporally global function fitting approaches in the context of the SMArtCARE registry, a prospective multicenter cohort study where longitudinal data on SMA patients' disease developments is collected during routine visits \citep{Pechmann2019}. As the registry has been set up relatively recently, there are few time points per patient with irregular timing and frequency of follow-up visits. 
In addition to an extensive baseline characterization, motoric ability is assessed with physiotherapeutic tests at follow-up visits. 

SMA is caused by a homozygous deletion in the survival motor neuron (SMN) 1 gene on chromosome 5, which is crucial for normal motor neuron function \citep{Pechmann2019}. As a result of SMN deficiency, muscles do not receive signals from motor neurons, leading to atrophy, i.e., muscle degeneration. Treatment is designed to increase production of the SMN protein to maintain functional motor neurons. These underlying disease processes cannot be observed directly, but are implicitly reflected in the motor function assessments, thus motivating modeling of latent dynamics that drive the observed measurements. 
We seek to identify these latent disease dynamics from the observed motor function using a VAE-based dimension reduction and model them conditional on patients' characteristics such as age, treatment and the extent of the SMN deficiency, which can be informative of the disease dynamics.

\begin{figure}
    \begin{center}
        \includegraphics[width=\textwidth]{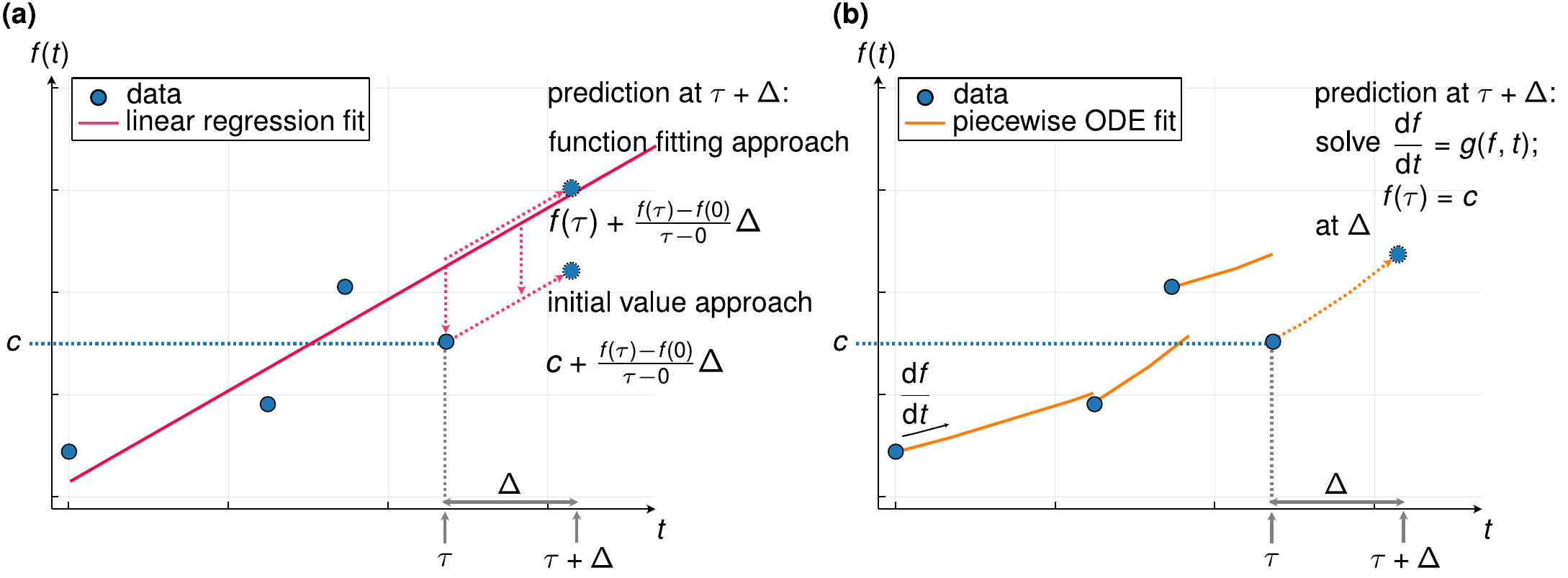}
        \vspace{-1.5cm}
    \end{center}
    \caption{Different perspectives on modeling dynamics: Regression-based function fitting approaches provide global fits of observed values on an absolute level (blue circles) (a), where the average fit (red solid line) is extrapolated (top red dotted line) to make predictions for new time points. In contrast, ODE-based approaches which use observations as initial values provide local fits by specifying the rate of change relative to the current status, which is used as the initial value for starting the ODE (b). Applying a similar perspective to a regression approach corresponds to shifting the regression fit to the current value and using that as initial value for starting the extrapolation (Panel (a), bottom red dotted line).}
    \label{fig:introfigure}
    \vspace{-0.5cm}
\end{figure}

For more closely analyzing the corresponding modeling challenge, we consider the observed time series of a hypothetical individual patient. In this setting, fitting a regression model corresponds to estimating a global function that is the best fit through the observed points on average, e.g., by minimizing squared distances (Figure \ref{fig:introfigure}a). In contrast, an ODE model describes the rate of change relative to the current observed value as an explicit mechanistic function (denoted as $g$ in Figure \ref{fig:introfigure}b), i.e., locally in time. For predictions in subsequent time intervals, the ODE is solved using the last observed value as initial condition (Figures \ref{fig:introfigure}b), while the average fit of all previous observations on an absolute level is used in regression-based function fitting (Figure \ref{fig:introfigure}a, function fitting approach). 

In this simplified setting, the regression approach can also be turned into an initial value approach by shifting the globally fitted function to the last observed value, which can then be used as starting point for extrapolation (Figure \ref{fig:introfigure}a, initial value approach). 
For example, in Figure \ref{fig:introfigure}, the slope equation of the regression is an approximation of the derivative of $f$ modeled by the differential equation for $\Delta \to 0$, i.e., in the limit both models are equivalent. Yet, regression coefficients do not translate to parameters of a local rate of change in general, but represent a conceptually distinct global function fitting approach. 

Global function fitting considers the observed values as measurements with noise, which is to be averaged out, within or often also across individuals, before extrapolating. For example, measuring the distance a SMA patient can walk within a given time frame will likely be subject to variability due to the patients' daily form and motivation. Using a global average for prediction in a regression-based approach then improves robustness to such random variations.  
In contrast, the last observation can be considered as the most accurate approximation of the true value of interest in ODE modeling. For example, when a child acquired a new motoric skill, extrapolating the future development based on having that skill might yield more meaningful predictions than averaging over all past time points when the child has not yet had that skill. 
In addition, the clinical practitioner might be more interested in anticipated relative changes, which are more directly reflected in ODE models, instead of the absolute levels of observations.
A typical function fitting approach cannot account for external changes that are not described by the model, whereas these can be taken into account by restarting the ODE based on the last observed value when a new observation is made. 
However, this comes at the price of a higher vulnerability to noise, illustrated by the jumps in Figure \ref{fig:introfigure}b, as opposed to a smooth global regression function. 

Another difference becomes apparent when assuming that model parameters of ODEs can be obtained from some other source instead of from direct function fitting. As function fitting for regression models is performed by minimizing the average distances of the absolute levels of observations over multiple time points and maybe even across multiple individuals, model fits will become more robust for longer observed time series, and less reliable when only a few data points are available. For example, this is problematic in the early phase of a clinical cohort study, such as the SMArtCARE registry, when only a small number follow-up times is available. 
In contrast, when assuming that ODE parameters can be obtained from external information such as baseline characteristics, the ODE is completely specified when just one initial observation is given as starting value, and can be used to predict subsequent developments. For example, patients' baseline characteristics could potentially be informative about subsequent individual disease progression in our SMA application, and specifically these connections might be of interest to biomedical researchers. While such an approach of determining model parameters based on baseline characteristics could, in principle, be developed for global regression models, this would imply that also the absolute level of the values could be predicted based on baseline characteristics, whereas in ODEs, they only need to inform a model for individual changes. 

Spelling out such conceptual differences can help to choose between the different approaches, 
depending on the application scenario and question to be answered. For example, the aim in our application is to anticipate an individual child's future motoric development based on present motoric skills, which favors the proposed ODE approach. 

\section{Methods}
\label{sec:methods}

\subsection{Statistical modeling with ODEs and many initial values}
\label{sec:multipleODEs}

Our proposed approach for estimating dynamics is based on solving a linear ODE system multiple times, using each value observed in the course of time as the initial condition. The resulting individual solutions are combined into an inverse-variance weighted average using time-dependent weights, to obtain an unbiased minimum variance estimator of the true underlying dynamics.
We assume a low-dimensional time series of observations $\bz_{t_0}, \dots, \bz_{t_K}$ with $\bz_t \in \R^d$ for all $t=t_0,\dots, t_K$, which can be thought of as noisy measurements of a true, deterministic underlying process $\bmu: \R_+ \to \R^d, t \mapsto \bmu(t)$,
i.e., $\bz_t = \bmu(t) + \beps(t)$, where $\beps(t)$ is a sample path of a stochastic process $(E_t)_{t\in \R_+}$ with $E_t \sim \mathcal{N}(0,\bfsigma^2_t)$ for all $t$. For simplicity, we assume that measurement errors are independent of time, i.e., $E_t \equiv E \sim \mathcal{N}(0,\bfsigma^2)$ for all $t\in \R_+$ for some fixed $\bfsigma^2 \in \R^d$. Thus, we can consider $(\bz_t)_{t=t_0,\dots, t_K}$ as a sample path of a stochastic process $(Z_t)_{t=t_0,\dots, t_K}$ with finite state space, where $Z_t \sim \mathcal{N}(\bmu(t), \bfsigma^2)$ for all $t$. 
We assume that the underlying dynamics can be described by a linear ODE system with constant coefficients, and consider for each $k = 0, \dots, K$ the initial value problem
\begin{align}
    \begin{split}
    \label{eq:ivp}
        \frac{d}{dt} \bmu(t) &= A\cdot \bmu(t) + \bc; \\
        \bmu(t_k) &= \bz_{t_k},
    \end{split}
\end{align}
where the $k$-th observation is used as the initial condition. 
For $A=0$, the solution to \eqref{eq:ivp} is given by $\bmu(t) = \bc\cdot t + \bz_{t_k}$.
For $A \neq 0$, the solution can be computed analytically (see Supplementary Section~1 for a detailed derivation):
    \begin{align}
        \label{eq:analyticalsolution}
        \bmu(t) =\exp(A(t- t_k))\cdot(A^{-1}\bc + \bz_{t_k}) - A^{-1}\bc.
    \end{align}

Note that for non-constant coefficients, an analytical solution can be obtained analogously but requires numerical evaluation of integral terms that often do not have closed-form solutions. 
Slightly overloading notation, we define the solutions of Equation \eqref{eq:ivp} for $A\neq 0$ as $\bz_k(t, \bfeta) :=(A^{-1}\bc + \bz_{t_k}) \exp(A(t- t_k)) - A^{-1}\bc$ for each $k=0,\dots, K$, where $\bfeta = \lbrace A, \bc \rbrace$ and the initial value is given by $\bz_k$. The solutions are then combined to a time-dependent inverse-variance weighted average as an estimator for the true underlying dynamics:
\begin{align}
\label{eq:estimator}
    g(t, \bfeta) := \frac{\sum_{k=0}^K \mathrm{Var}\left[\bz_k(t, \bfeta)\right]^{-1} \bz_k(t, \bfeta)}{\sum_{k=0}^K \mathrm{Var}\left[\bz_k(t, \bfeta)\right]^{-1}}.
\end{align}

As $\mathbb{E}\left[\bz_k(t, \bfeta)\right] = \bmu(t_k)$ for all $k=0, \dots, K$, 
the estimator $g(t, \bfeta)$ is unbiased. Further, it has minimum variance among all weighted average estimators \citep{Shahar2017}. 
For calculating the weights, we have 
\begin{align}
    \begin{split}
        \mathrm{Var}\left[\bz_k(t, \bfeta)\right] &= \mathrm{Var}\left[\exp(A(t-t_k)) \cdot (A^{-1}\bc + \bz_{t_k}) - A^{-1}\bc\right]\\ 
        &= \mathrm{Var}\left[\exp(A(t-t_k) \cdot \bz_{t_k} \right] \\ 
        &= \exp(2A(t-t_k)) \cdot \mathrm{Var}\left[\bz_{t_k} \right] \\ 
        &= \exp(2A(t-t_k)) \cdot \bfsigma^2,
    \end{split}
\end{align}
which we can insert in \eqref{eq:estimator} to obtain an explicit closed-form expression. When $\sigma^2$ is unknown, we estimate $\mathrm{Var}\left[\bz_k(t, \bfeta)\right]$ by calculating the sample variance of the  ODE solutions $\bz_j(t, \bfeta)$ for all $j$ with $t_k < t_j < t$, as described below in Section~\ref{sec:ourmodel}.

\subsection{Modeling dynamics in a latent representation}
\label{sec:dynamicsinlatent}

When there is a large number of variables that are observed over the course of time, often a model for an underlying lower-dimensional process is sought that drives the observed measurements, instead of separately modeling the dynamics of each observed variable. For example, in our SMA application, the underlying disease dynamics are driven by degradation of motor neurons causing muscle degeneration, which cannot be observed directly but is implicitly reflected in various items of motor function tests. 
To model such latent processes, we suggest to use neural networks for learning a low-dimensional representation of observed measurements, and integrating our statistical approach for dynamic modeling with ODEs into such a latent representation via simultaneous fitting of the dynamic model and the neural networks.

\subsubsection{Using VAEs for dimension reduction}
\label{sec:VAEmodel}

For dimension reduction, we use a variational autoencoder (VAE), a generative deep learning model that infers a compressed, low-dimensional representation of the data using neural networks \citep{Kingma2014, Kingma2019}. 
Specifically, the latent space is defined as a low-dimensional random variable $\bcz$ with prior distribution $P^{\bcz}$. Two distinctly parameterized neural networks, called the encoder and the decoder, map an observation $\bx \in \R^{p}$ to the latent space and back to data space. The parameters of the encoder and decoder are jointly optimized to infer a low-dimensional representation, based on which the original data can be well reconstructed.
Formally, the encoder and decoder parameterize the conditional distributions $q_{\bcz \mid \bx}(\cdot,\bfphi)$ and $p_{\bcx \mid \bz}(\cdot,\btheta)$, where we abbreviate conditional distributions and densities by writing, e.g., $p_{\bcx\mid \bz}(\bx, \btheta)$ for $p_{\bcx \mid \bcz =\bz}(\bx, \btheta)$. The model is trained, i.e., the parameters $\btheta$ and $\bfphi$ of the encoder and decoder networks are optimized, by maximizing the evidence lower bound (ELBO), a lower bound on the data likelihood $p_{\bcx}$, derived based on variational inference \citep{Blei2017}:
    \begin{equation*}
        \mathrm{ELBO}(\bx, \bfphi, \btheta) = E_{q_{\bcz \mid \bx}(\cdot, \bfphi)}[\log(p_{\bcx \mid \bz}(\bx, \btheta))] - D_{\mathrm{KL}}(q_{\bcz \mid \bx}(\cdot, \bfphi)\Vert p_{\bcz}).    
    \end{equation*}

It can be shown that maximizing the ELBO is equivalent to minimizing the Kullback-Leibler divergence between the true but intractable posterior $p_{\bcz \mid \bx}(\cdot,\btheta)$ and its approximation by a member of a parametric variational family $\lbrace q_{\bcz \mid \bx}(\cdot,\bfphi) \rbrace$, typically assumed as Gaussian with diagonal covariance matrix \citep{Blei2017, Kingma2014, Kingma2019}. 
The first term of the ELBO can be interpreted as a reconstruction error, while the second term acts as a regularizer that encourages densities close to the standard normal prior.

Training the VAE then corresponds to maximizing the ELBO as a function of the parameters $\btheta$ and $\bfphi$ of the encoder and decoder, specifying the conditional distributions, which yields both an approximate maximum likelihood estimate for $\btheta$ and an optimal variational density $q_{\bcz \mid \bx}(\cdot, \bfphi)$. Such optimization is typically performed by stochastic gradient descent \citep{Kingma2015}, where the so-called reparameterization trick is used to obtain gradients of the ELBO w.r.t. the variational parameters $\bfphi$ \citep{Kingma2014, Kingma2019}. 

\subsubsection{Incorporating dynamics into the loss function}
\label{sec:ourmodel}

Building on an approach that we have proposed previously \citep{HacHarPfa2022}, we integrate the dynamic model described in Section \ref{sec:multipleODEs} in the VAE latent space. 
We use observed time series of patients' measurements as input to the model, represented as a matrix $\bx_i$ of $T_i+1$ measurements of $p$ variables for individual $i$, where $t_0$ is the common baseline time point, and $t_{1}^i, \dots, t_{T_i}^i$ are individual-specific subsequent measurement time points. In our application setting from clinical cohort studies, typically a more extensive patient characterization at the baseline time point is available, such as age at symptom onset or treatment start, which we assume to be informative of intra-individual differences in underlying disease dynamics.

We specify the variational posterior as multivariate Gaussian with diagonal covariance matrix, parameterized by the VAE encoder. Specifically, the encoder maps an observed time series column-wise to the posterior mean $\bmu_i= (\bmu_i^{t_0}, \bmu_i^{t_1^i}, \dots,  \bmu_i^{t_{T_i}^i}) \in \R^{m\times T_i+1}$ and standard deviation $\bfsigma_i = (\bfsigma_i^{t_0}, \bfsigma_i^{t_1^i}, \dots,  \bfsigma_i^{t_{T_i}^i}) \in \R^{m\times T_i+1}$. 
We assume that the dynamics of the latent posterior mean are governed by a linear ODE system and use the approach described in Section \ref{sec:multipleODEs} to calculate the estimator from Equation \eqref{eq:estimator} and define 

\begin{equation}\label{eq:ODE_estimator}
    \widetilde{\bmu_i}(t, \bfeta_i) := \frac{\sum_{k=0}^{T_i} \mathrm{Var}[\widetilde{\bmu}_{i,k}(t, \bfeta_i)]^{-1} \widetilde{\bmu}_{i,k}(t, \bfeta_i)}{\sum_{k=0}^{T_i} \mathrm{Var}[\widetilde{\bmu}_{i,k}(t, \bfeta_i)]^{-1}}.    
\end{equation}

Here $\widetilde{\bmu}_{i,k}(t, \bfeta_i)$ is the analytical ODE solution obtained according to Equation \eqref{eq:analyticalsolution}, using the encoded value $\bmu_i^{t_k^i}$ from the $k$-th time point of the $i$-th individual as initial value. 

To estimate the unknown variance $\mathrm{Var}[\widetilde{\bmu}_{i,k}(t, \bfeta_i)]$ at a time point $t$ of the ODE solution with initial value $\bmu_{i}^{t_k^i}$, we use all encoded values $\bmu_{i}^{t_j^i}$ for $t_k^i < t_j^i < t$, i.e., from time points between the current initial time point ${t_k^i}$ and the time point $t$ of interest, and calculate the sample variance of the corresponding ODE solutions $\widetilde{\bmu}_{i,j}(t, \bfeta_i)$ for all $j$ with $t_k^i < t_j^i < t$.

To obtain personalized dynamics conditional on baseline information, we use an additional neural network to map each patient's baseline variables to a set of individual ODE parameters $\bfeta_i$. 
We subsequently use the estimator $\widetilde{\bmu_i}(t, \bfeta_i)$ of the underlying dynamics for sampling $\bz_i^{t_k^i} \sim \mathcal{N}(\widetilde{\bmu_i}(t_k^i, \bfeta_i), \bfsigma_i^{t_k^i})$ for $k=0, \dots, T_i$, and passing it to the VAE decoder to obtain a reconstructed time series $\widehat{\bx}_i$. 

As the parameters $\btheta$ and $\bfphi$ are assumed to define the common distributions which underlie all individual observations, these parameters and hence also the weights of the VAE encoder and decoder are shared for all individuals, as well as the weights of the additional baseline network. We thus optimize a single VAE and baseline neural network. Yet, as each patient's individual time-dependent and baseline variables are used to obtain their latent representation and ODE parameters, the loss function is calculated for each individual. 
For jointly optimizing the VAE and the neural network for the ODE parameters, we iteratively maximize the ELBO, using $\widetilde{\bmu_i}(t, \bfeta_i)$ as the latent posterior mean. We add the squared Euclidean distance between the posterior mean $\bmu_i$, obtained directly from the encoded time series, and the posterior mean constrained to smooth dynamics $\widetilde{\bmu_i}(t, \bfeta_i)$ to encourage consistency of the latent representation before and after solving the ODEs. To regularize the inverse-variance weights of the ODE solution estimator, we further add the log-differences between the sample variances $s^2$ of all solutions $\widetilde{\bmu}_{i,k}(t, \bfeta_i), k=0,\dots, T_i$, and of all encoded values $\bmu_i$, summed across all observed time points, where an offset of $0.1$ is added before taking the logarithm to avoid values near zero. 
The final loss function for a single individual $i$ is given by 
\begin{equation} \label{eq:finalELBO}
	\begin{split}
		\mathcal{L}(\bx_i, \bfeta_i, \btheta, \bfphi) 
		&= D_{\mathrm{KL}}(\widetilde{q}_{\bcz \mid \bx_i}(\cdot, \bfeta_i, \bfphi) \Vert p_{\bcz}(\cdot,\btheta)) - \mathrm{E}_{\widetilde{q}}[\log(p_{\bcx \mid \bz_i}(\bx_i,\btheta))] \\
		&+ \alpha \Vert\bmu_i - \widetilde{\bmu_i}\Vert_2^2 \\
        &+ \beta \sum_{k=0}^{T_i} \log(s^2((\widetilde{\bmu}_{i,j}(t_k, \bfeta_i))_{j=0,\dots, T_i}) + 0.1) - \log(s^2((\bmu_i^{t_j})_{{j=0,\dots, T_i}}) + 0.1)
	\end{split}
\end{equation}
where $\alpha, \beta \in [0,1]$ are hyperparameters balancing the loss components.

We optimize the joint loss function from Equation \eqref{eq:finalELBO}, which simultaneously incorporates all model components, i.e., the dynamic model, the VAE for dimension reduction, and the neural network for obtaining the ODE parameters from baseline characteristics, by stochastic gradient descent. In our implementation, we perform a gradient update step based on the ELBO of each single individual, i.e., we use batches of size 1. When using larger batch sizes, we could average the ELBOs of all individuals in the batch before performing an update step. 

We use automatic differentiation \citep{Baydin2017} to simultaneously obtain gradients with respect to all parameters, including the VAE encoder and decoder parameters $\btheta, \bfphi$ and the individual-specific dynamic model parameters $\bfeta_i$. This can be realized efficiently using the Zygote.jl package \citep{Innes2019_Zygote_dP} in the Julia programming language \citep{Bezanson2017}, a flexible framework for automatic differentiation that allows to implement the model and loss function with only minimal code adaptation for automatically obtaining gradients. 
The model is implemented as a publicly available Julia package  (\url{https://github.com/maren-ha/LatentDynamics.jl}), including Jupyter notebooks to illustrate the approach. 
Further implementation details can be found in Supplementary Section~2. 

\section{Evaluation}
\label{sec:evaluation}

\subsection{Simulation}
\label{sec:simulation}

We empirically evaluate our approach in a simulation design and in an application with the SMArtCARE rare disease registry on SMA patients' development of motoric ability. 
A general challenge for evaluation is that the latent representation is invariant to affine linear transformations such, as shifting, scaling and rotation, as these can be reversed in the decoder, such that scaling in latent space is arbitrary. In this setting, our approach should, e.g., distinguish patterns with different monotonicity behavior as a minimum requirement. For example, in SMA there might be two underlying processes corresponding to, e.g., motor neuron degradation and muscle function, and a model should be able to distinguish between groups of patients with different development trends. 

To investigate this property in a simulation study, we adapt our previous design from \citet{HacHarPfa2022}. There, we defined two groups of individuals with distinct underlying development patterns, corresponding to two distinct sets of ODE parameters, based on a homogeneous two-dimensional linear ODE system, i.e., with four unknown parameters. The two ODE systems are given by

\begin{minipage}{.45\linewidth}
	\begin{equation*}
	\begin{split}
		\frac{d}{dt}\begin{pmatrix} u_1 \\ u_2 \end{pmatrix}(t) &= \begin{pmatrix} -0.2 & 0.1 \\ -0.1 & 0.1 \end{pmatrix} \begin{pmatrix} u_1 \\ u_2 \end{pmatrix}(t); \\
		\begin{pmatrix} u_1 \\ u_2 \end{pmatrix}(0) &= \begin{pmatrix} 3 \\ 1 \end{pmatrix}
	\end{split}
	\end{equation*}
\end{minipage}\begin{minipage}{.55\linewidth}
	\begin{equation*}
	\begin{split}
		\frac{d}{dt}\begin{pmatrix} u_1 \\ u_2 \end{pmatrix}(t) &= \begin{pmatrix} -0.2 & -0.1 \\ 0.1 & -0.2 \end{pmatrix} \begin{pmatrix} u_1 \\ u_2 \end{pmatrix}(t); \\
		\begin{pmatrix} u_1 \\ u_2 \end{pmatrix}(0) &= \begin{pmatrix} 3 \\ 1 \end{pmatrix}.
	\end{split}
	\end{equation*}
\end{minipage}
\vspace{0.5cm}

We simulated $n=100$ individuals split equally into the two groups. For each individual $i=1, \dots, n$, we sampled a random number $T_i$ of between $1$ and $8$ follow-up observations after the common baseline time point $t_0$ at random time points $t_1^i, \dots, t_{T_i}^i$, sampled uniformly from $[1.5,10]$. At each time point $t_k^i$, we simulated measurements of $p=10$ variables by adding a variable-specific measurement error $\delta_{j}^{t_k^i} \sim \mathcal{N}(0,\sigma_{\mathrm{var}}^2)$ for $j=1,\dots, p$ and an individual-specific measurement error $\varepsilon_{i,j}^{t_k^i} \sim \mathcal{N}(0,\sigma_{\mathrm{ind}}^2)$ for $i=1,\dots, n$, $j=1,\dots, p$ to the true value of the ODE solution. 
For an individual $i$, we then obtained simulated observations $\bx_i \in \mathbb{R}^{p \times T_i + 1}$ by defining 
\begin{equation*}
    x_{i,j}^{t_k} := u_1(t_k) + \delta_{j}^{t_k} + \varepsilon_{i,j}^{t_k} \quad \text{for } k=0, \dots, T_i+1 \text{ and } j = 1, \dots, 5,
\end{equation*}
and 
\begin{equation*}
x_{i,j}^{t_k} := u_2(t_k) + \delta_{j}^{t_k} + \varepsilon_{i,j}^{t_k} \quad \text{for } k=0, \dots, T_i+1 \text{ and } j = 6, \dots, 10.
\end{equation*}

We additionally simulated baseline variables by sampling with variance $\sigma_{\mathrm{info}}^2$ from the simulated individual's true ODE parameters to obtain $10$ informative baseline variables and add $40$ noise variables sampled from a centered Gaussian distribution with variance $\sigma_{\mathrm{noise}}^2$, such that we ended up with a total of $q = 50$ baseline variables.
We set $\sigma_{\mathrm{ind}} = 0.5$ and $\sigma_{\mathrm{var}} = \sigma_{\mathrm{info}} = \sigma_{\mathrm{noise}} = 0.1$. 

On this data, we trained the model described in the previous section and extracted the learnt trajectories (see Supplementary Section~2 for training details). Hyperparameters, including the weights to balance the loss components, were chosen based on monitoring convergence of the loss function and visualizing fitted latent trajectories without systematic hyperparameter tuning. 

\begin{figure}
    \begin{center}
        \includegraphics[width=0.8\textwidth]{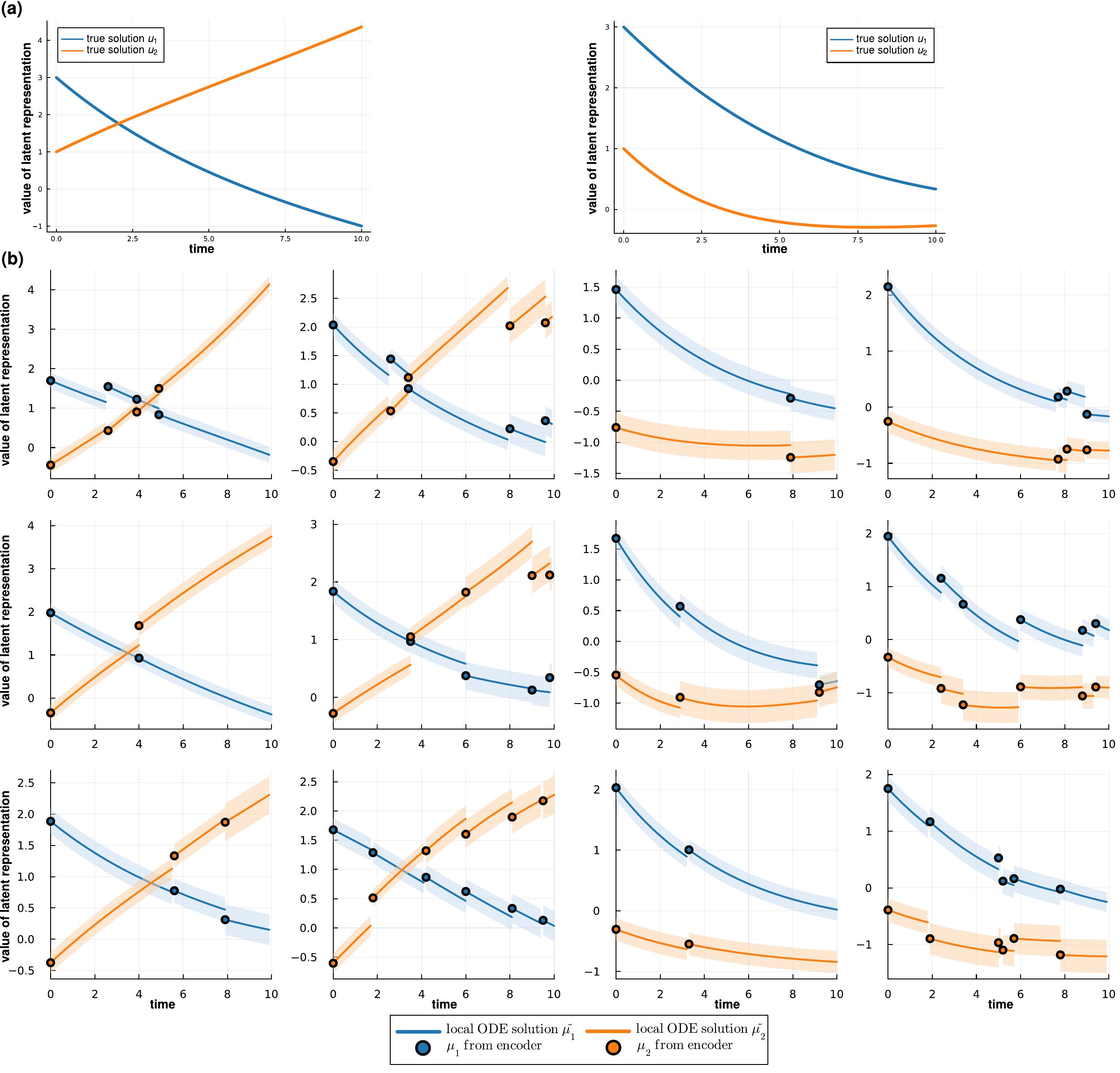}
    \end{center}
    \caption{(a) Ground-truth trajectories for two distinct development patterns based on which data was simulated. (b) Exemplary fits of simulated individuals from both groups. One panel corresponds to one individual, the $x$ axis shows the time and the $y$ axis the values of the two dimensions of the latent representation. Circles depict the mean of the latent variables, as obtained directly from the encoder, while solid lines indicate the ODE solutions, with parameters obtained from the individual's baseline variables, using the value at the last observed time point as initial condition and solved until the subsequent time point. Colored bands show the range of one standard deviation of the latent variable around its mean. The two leftmost columns show $6$ individuals from the first group, corresponding to the left ground-truth trajectory in (a), while the two rightmost columns show $6$ individuals from the second group, corresponding to the pattern shown on the right in (a).}
    \label{fig:sim_results}
\end{figure}

The focus of our approach is to locally predict changes in latent dynamics, given the last measurement.  
Correspondingly, we visualize the learnt trajectories by starting at the learnt latent representation of each observed time point consecutively and solving an ODE with the learnt parameters until the next time point, thus reflecting our aim of predicting changes in the immediate future based on the current status until a new measurement becomes available. In Figure \ref{fig:sim_results}, we show the true underlying ODE solutions of both groups in Panel (a) and the fitted latent trajectories (solid lines) based on the mean of the latent representation for $12$ exemplary simulated individuals in Panel (b). The colored bands around the trajectory correspond to the learnt standard deviation of the latent representation. The results show that the approach allows for recovering group-specific underlying dynamic patterns, as the fitted trajectories match the group-specific true ODE solution shown in Panel (a).  The local predictions at the subsequent time point mostly fall within the range of one standard deviation (colored bands) of the next encoded observation (filled circles). 
Note that the fitted trajectories sometimes appear shifted or downscaled (in particular in the second dimension), due to the model freedom in structuring its latent representation mentioned above and the $\mathcal{N}(0,1)$ prior on the latent variable.

To verify robustness of the approach, we have repeated the data simulation and model fitting process $20$ times, where for each dataset resampling we have also resampled the random parameter initialization of the neural network components. We have calculated relative change in prediction error with respect to the regression approach as a baseline, with a mean relative improvement of $62,11$\%, a minimum relative improvement of $27.07$\% and a maximum relative improvement of $74.11$\% over the baseline. 

Further, we conducted a sensitivity analysis by repeating the simulation with different levels of noise in the data generation and different numbers of simulated individuals and time points per individual. The results show that, as expected, the performance improves with larger numbers of observations and time points and a lower level of noise, while it is relatively robust with respect to the number of time points per individual. Details are reported in Supplementary Section~3 and in Supplementary Figure~1.

In addition, we wanted to empirically verify that our approach indeed coincides with a linear regression fit when the underlying dynamics are simple, i.e., it defaults to a least-squares fit (see Section~\ref{sec:illustration}). We thus considered a simpler simulation scenario with a constant two-dimensional ODE-system, i.e., resulting in linear fits, where estimating the rate parameter of the ODE and the initial condition is equivalent to fitting a simple linear regression with time as dependent variable separately for each latent dimension. In this setting, local ODE fits from our model indeed closely match a global linear regression fit of the encoded time series (see Supplementary Figures~2 and 4. 
To fit this global regression model, the complete time series has to be observed, while ODE solutions are obtained using baseline information and the value at the last observed time point only. If only this limited information is available and extrapolations have to be calculated based on previously observed values only, the regression approach performs worse, as reflected by a substantial drop in prediction performance compared to the ODE approach, both in latent space and on the level of the reconstructed items (see Supplementary Section~4).

\subsection{Rare disease registry application}
\label{sec:smartcare}

To illustrate our approach with real data, we use data from the SMArtCARE rare disease registry. Specifically, we selected patients treated with Nusinersen \citep{Schorling2020} who completed the Revised Upper Limb Module (RULM) test \citep{Mazzone2017}. This test comprises $20$ items with a maximum sum score of $44$ to evaluate motor function of upper limbs and can be conducted for all patients older than two years of age with the ability to sit in a wheelchair \citep{Pechmann2019}. 
We use all test items as time-dependent variables and all available baseline information, including, e.g., SMA subtype, age at symptom onset and first treatment, and genetic test results. 
We applied three filtering steps to the dataset based on clinical considerations.
First, we excluded $63$ patients with fewer than two observation time points, as meaningful trajectory estimation is not feasible for such cases.
Second, we filtered out $176$ patients with RULM sum score variance smaller than $1$, i.e., with nearly constant trajectories across all items, as the prediction of changes in dynamics is of main clinical interest. 
Third, for each patient we have removed outlier time points where the difference in RULM sum score to the previous time point exceeded two times the interquartile range of all sum score differences between adjacent time points, as clinicians assume relatively smooth disease trajectories, where abrupt jumps are typically disregarded as outliers.
This resulted in a dataset of $399$ patients with observations at between $2$ and $13$ time points (median $7$ time points), corresponding to in total $2797$ observations of $20$ time-dependent variables, in addition to $24$ baseline variables. Integer-valued test items were rescaled and a logit transformation was applied to account for the Gaussian generative distribution parameterized by the VAE decoder.
\begin{figure}
    \begin{center}
        \includegraphics[width=\textwidth]{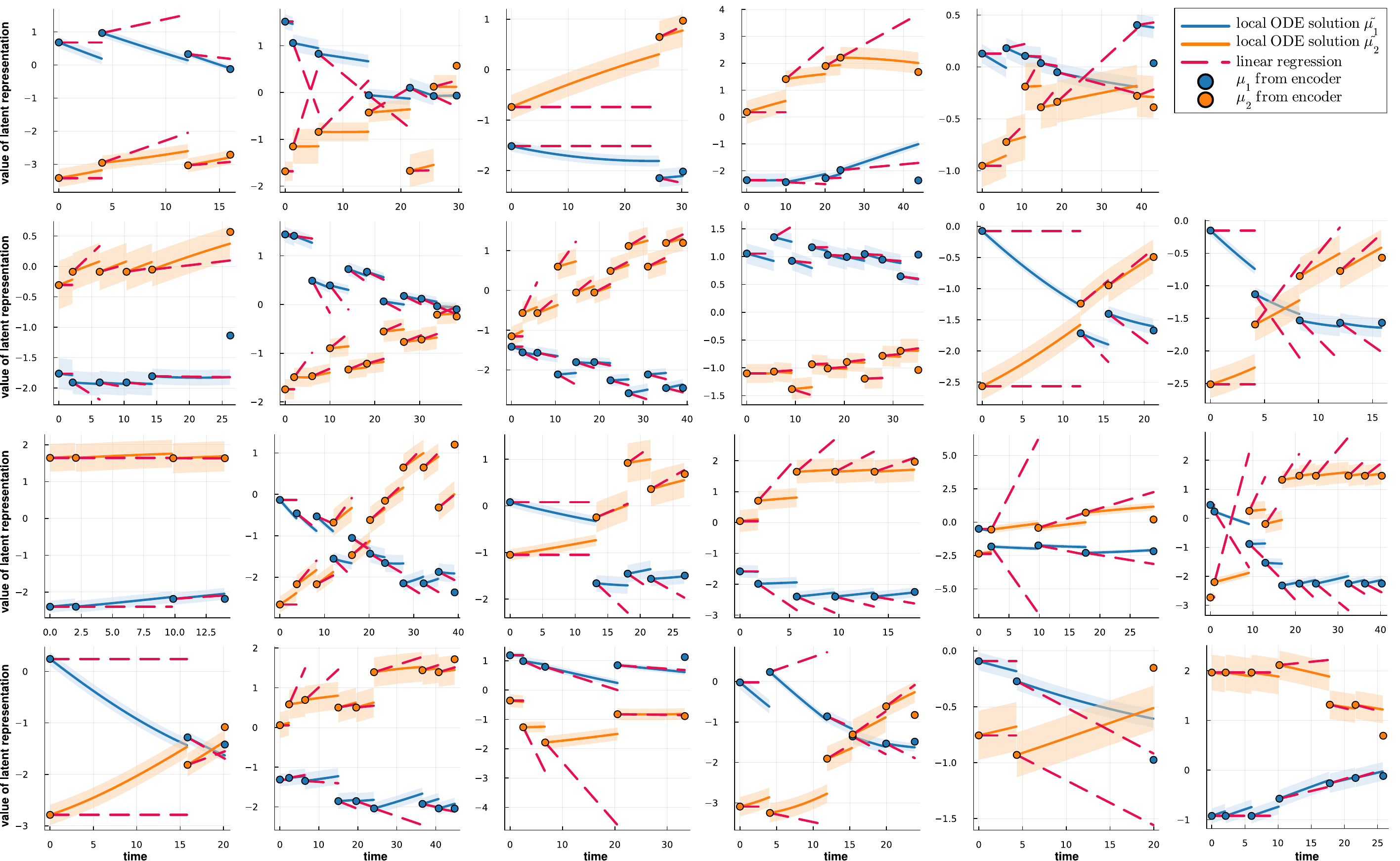}
    \end{center}
    \caption{Exemplary fits from SMArtCARE patients. Each panel shows the learnt latent space trajectories of one patient, where the $x$ axis shows the time in months and the $y$ axis the values of the two dimensions of the latent representation. Circles depict the latent variable means obtained directly from the VAE encoder, solid lines depict the fitted ODE solution using the last observed value as initial condition, and solved until the subsequent observation. Colored bands show a range of one standard deviation of the latent variables around their mean. Least squares regression fits of all previously observed latent values, shifted to the current observed time point and extrapolated until the subsequent observation, are shown in pink.}
    \label{fig:smartcare}
\end{figure}
In Figure \ref{fig:smartcare}, we show fitted latent trajectories from a two-dimensional ODE system for the latent space for exemplary SMArtCARE patients. Analogous to Figure \ref{fig:sim_results}, we display the ODE solutions for the posterior mean of the latent representation with the learnt parameters, using the first encoded value as the starting point and solving until the subsequent observation time point, where the ODE is restarted with the new value as initial condition. 

For comparison, we computed least squares regression fit of all previously observed encoded values at each observed time point and shifted the trajectory to start at the current observation and extrapolated until the next observation (red lines), as described above. 
Especially at the beginning of the time series, the extrapolated values are often far off from the observed value. For example, for the first observed time point, the current value can only be carried forwards in time, i.e., predicting a constant trajectory. At later time points, changes in monotony behavior lead to significant prediction errors (e.g., in the third panel from left in the bottom row), while in general predictions tend to get more accurate at later time points, when more information has already been accumulated (e.g., as in the leftmost panel in the second row). 

This illustrates the conceptual difference in perspective to the ODE-based approach, where the accuracy of the prediction does not depend on the amount of previously observed values, as it relies on information from baseline characteristics and acts locally as opposed to global models. Here, predicted values mostly fall within the range of one standard deviation of the mean trajectory (colored bands). Our proposed approach is able to capture individual-specific patterns as reflected by individuals with various different trends and both very dynamic and more constant patterns. 
Joint optimization helps to regularize the latent representation, as the ODE model imposes a smoothness constraint, thus encouraging a representation where encoded values fall on a smooth trajectory. While this is effective in most cases, there are still some outlier time points (e.g., in the second panel in the top row) or large jumps that cannot be captured by a smooth model. 

In addition to the conceptual comparison with the shifted regression approach, we also compared our approach with several alternative approaches for multivariate longitudinal modeling. Specifically, we have developed additional versions of the regression approach, namely using the model predictions directly without applying a shift to the current value, incorporating the baseline variables used for obtaining individual ODE parameters, and including time with a quadratic effect to allow for more flexibility. Additionally, we have fitted continuous-time autoregressive models \citep{Wang2013} to each individual time series, and applied the functional PCA approach of \citet{Nolan2023} for identifying a smooth joint trajectory with individual deviations. 

The different regression approaches and the autoregressive model were fitted on the latent representation, while the functional PCA approach was fitted on the original data. For comparing with our approach and the shifted regression baseline, we calculated prediction errors in latent space and on item-level decoder reconstructions of the model predictions. 

To incorporate baseline information analogous to our ODE-based approach, we adapted the regression approach as follows: First, we fit a linear model of the complete latent time series, separately for each individual, analogous to the setting of our ODE approach, where also an individual's complete latent time series is available during model training. Next, we collected the slopes of these fits for all individuals, corresponding to the ODE parameters, and fit a linear model to predict the slopes with individuals' baseline variables, analogous to the optimization of our additional baseline network. We finally used the predicted slopes based on this model to predict the subsequent time point, starting at each latent value sequentially. 

For the functional PCA model, we used the default hyperparameters as provided in the original authors' implementations. For the autoregressive models, we adjusted the ``scale'' value in the model fitting function to a smaller number, which led to better results on our data. 
We used first-order (AR(1)) models, as our relatively short time series (with a median of 8 time points) make it challenging to reliably estimate more parameters, as would be required for higher-order autoregressive models. Even with AR(1) models, we noted several cases of non-convergence. For these, we tried re-fitting the models with a smaller ``scale'' value, which led to convergence in some but not all cases. The remaining cases of non-convergence were excluded from the calculation of the mean squared errors. 
The AR(1) models had to be fitted separately for each latent dimension for each individual, as the approach did not provide a multivariate option.

The results are summarized in Tables~\ref{tab:predictions_modelcomparisons_latent} and~\ref{tab:predictions_modelcomparisons_reconstructed}, showing that the ODE-based approach consistently provides lower prediction errors compared to the alternative approaches. In particular, the closest competitor is the adaptation of the regression with the baseline variables, which is conceptually most similar to our ODE approach.

\begin{table}[ht]
    \centering
    \begin{tabular}{l|c}
        Model & Prediction in latent space  \\ \hline
        ODE & 1.386  \\
        Regression with baseline variables & 1.485 \\
        Regression with shift & 3.523  \\
        Regression without shift & 3.571 \\
        Autoregressive model (AR(1)) & 6.423 \\
        Regression with quadratic time effect and shift & 137.569 \\
        Regression with quadratic time effect and no shift & 137.633 \\
    \end{tabular}
    \caption{Mean squared prediction error in latent space, averaged across all time points and all SMArtCARE patients, when using our ODE-based approach, different versions of a regression approach, or a first-order autoregressive model.}
    \label{tab:predictions_modelcomparisons_latent}
\end{table}

\begin{table}[ht]
    \centering
    \begin{tabular}{l|c}
        Model & Prediction on reconstructed values \\ \hline
        ODE & 5.181 \\
        fPCA &  19.611 \\
        Regression with shift &  20.890 \\
    \end{tabular}
    \caption{Mean squared prediction error on the level of the reconstructed items, averaged across all time points and all SMArtCARE patients, when using our ODE-based approach, a simple shifted regression approach, or a functional PCA model.}
    \label{tab:predictions_modelcomparisons_reconstructed}
\end{table}

\section{Discussion}
\label{sec:discussion}

While regression techniques are commonly used for modeling dynamic processes in statistics, ODE-based approaches dominate in the systems modeling community. We have shown that different approaches correspond to conceptually different perspectives on dynamic modeling and have highlighted several aspects that might be considered for deciding on a modeling approach in a given application scenario. In particular, our intention has been to illustrate how ideas can be combined across communities, as in adapting ODE-based modeling for longitudinal cohort data. Specifically, we were motivated by an SMA rare disease registry as a prototypical example in a clinical setting, where underlying disease dynamics are to be inferred from a larger number of observed variables and modeled in a latent space, using individual-specific models. There, a particular challenge is that observations are noisy, irregular in time, and heterogeneous. 

Regression approaches typically correspond to function fitting techniques, where the starting point for subsequent assessments is an extrapolation of an average fit of all previously observed data on an absolute level, within and often across individuals, whereas in ODE-based models, 
relative changes to a starting point are modeled. We have argued that this difference is not merely technical, but reflects different analysis intentions, relating to what is considered the closest approximation of an underlying truth. 
This is closely linked to the conceptualization of variability. Function fitting approaches implicitly or explicitly assume measurements as noisy, such that using an average increases robustness and is considered more accurate, whereas ODE approaches using the last observed value as starting point consider this last observed value as most representative and are thus more sensitive to noise. While challenging when dealing with noisy data, as frequently encountered in biomedical applications, predicting future developments conditional on a patient's current status locally in time with such an ODE approach is attractive, as it might more closely match the perspective of a clinician who seeks to predict immediate changes in the near future based on a patient's current status, rather than a global average. 
We have argued that the local perspective in ODE-based approaches further allows for modeling each patient's time series individually based on external information. This is attractive in particular when heterogeneity of data can be explained by baseline characteristics, and in settings where not many follow-up time points are available yet.

To nonetheless incorporate the advantages of statistical function fitting approaches, e.g., for dealing with noise, we have thus proposed a statistical approach based on ODEs, where we decrease dependence on the initial condition by combining multiple ODE solutions into an inverse-variance weighted unbiased estimator of the underlying dynamics. To allow for personalized trajectories, we have used patients' baseline characteristics to infer individual-specific ODE parameters. For modeling with a larger number of variables, we have combined our approach with dimension reduction by a VAE, which allowed for simultaneous fitting of all model components. This enables the use of ODEs also for noisier observations with multiple variables, such as in biomedical applications. 

In a simulation design and an application on SMA patient's motor function development data, we have shown that the approach allows for inferring individual-specific trajectories in the VAE latent space and for predicting subsequent observations. 
Notably, the approach provides personalized predictions on a patient's development immediately upon study entry, as their baseline measurements and first measurement of the time-dependent variables allow for fully specifying an individual-specific ODE system. 
Joint optimization of the VAE and the dynamic model, facilitated by differentiable programming, allows for adapting the latent representation to the dynamic model, which acts as a regularizing smoothness constraint. 

Yet, the latent representation is only identifiable up to an affine linear transformation, due to the encoder neural network structure. While the dynamic model parameters can be interpreted in the context of the latent representation, it might be challenging to link this interpretation to the observed data. 
For facilitating clinical insight, the latent representation could be constrained more strongly or combined with a post-hoc explainability technique. Alternatively, latent trajectories can be decoded back to observation space and the data-level trajectories can be queried for clinical patterns. 

More generally, interpreting the latent representation and choosing a suitable ODE system is challenging. In the present application, we used a simple linear ODE system with two components to reflect the underlying changes in motor neuron functionality and muscle fitness as potential driving forces, and also preferred a model with few parameters as the number of observations was rather small. Alternatively, specification of the ODE system could be addressed by a model selection strategy, e.g., by comparing prediction performance of models with different degrees of complexity in a stepwise approach. 

Similarly, uncertainty quantification in the latent representation remains challenging. While we reflect the standard deviation of the VAE's latent posterior in Figures~\ref{fig:sim_results} and \ref{fig:smartcare}, this likely underestimates the true uncertainty in the trajectories. Multiple sources of uncertainty, such as encoder variability, noise in the observations, and uncertainty in the dynamic model, are all entangled in the latent representation. Additionally, the variational approximation (Gaussian with diagonal covariance) might be too restrictive to capture the full posterior variability, and the model does not explicitly distinguish between epistemic and aleatoric uncertainty. Disentangling and quantifying the different sources of uncertainty therefore remains an important open problem for future work.

A further current limitation of the proposed approach is that time points of follow-up visits are assumed to be independent of the latent state. It might be interesting to relax this assumption in future work. Then, prediction uncertainty could be used as a criterion for determining when a patient should ideally be seen again.

In summary, the proposed approach provides a flexible modeling alternative for assessing individual-specific dynamics in longitudinal cohort data, as illustrated in a prototypical application scenario, and more generally exemplifies the benefits of integrating advantages of different modeling strategies across communities. 

\bigskip

\section*{Conflict of interest}

The authors declare that there are no conflicts of interest.

\section*{Data availability}

The data from the SMArtCARE registry cannot be shared due to ethical and legal restrictions. 
The simulated data, which mimics the structure and format of the real data, can be fully reproduced using the code provided in the public GitHub repository \url{https://github.com/maren-ha/LatentDynamics.jl}. 

\bigskip

\bibliographystyle{agsm}
\bibliography{references.bib}
\end{document}